\documentclass[aps,prfluids,twocolumn,twoside, a4paper,nofootinbib,groupedaddress,superscriptaddress,11pt]{revtex4-2}
\usepackage[caption=false]{subfig}     % required only when side-by-side / subfloats are used
\usepackage{amsmath}
\usepackage{amssymb}%% The amssymb package provides various useful mathematical symbols
\usepackage{graphicx} % Required for including pictures
\usepackage{ifpdf}\ifpdf % si pdftex
\usepackage[linktocpage,pdftex,hyperfigures]{hyperref}
\else
\usepackage{url}
\newcommand{\email}[1]{\rm{#1}}
\newcommand{\href}[2]{#2}%

\usepackage{hyperref}
\fi
\usepackage{siunitx}
\usepackage{amssymb}

\graphicspath{ {./figures/} }

\begin{document}

\title{Deterioration of water evaporation by impurity accumulation at the liquid-vapor interface during nucleate boiling}
\author{Corentin Le Houedec}
\affiliation{STMF, Universit\'e Paris-Saclay, CEA, 91191 Gif-sur-Yvette Cedex, France}
\author{Cassiano Tecchio}\email[Contact email: ]{cassiano.tecchio@cea.fr}
\affiliation{STMF, Universit\'e Paris-Saclay, CEA, 91191 Gif-sur-Yvette Cedex, France}
\author{Bo\v stjan Zajec}
\affiliation{STMF, Universit\'e Paris-Saclay, CEA, 91191 Gif-sur-Yvette Cedex, France}
\author{Pere Roca i Cabarrocas}
\affiliation{LPICM, CNRS, Ecole Polytechnique, Institut Polytechnique de Paris, 91120 Palaiseau, France}
\author{Pavel Bulkin}
\affiliation{LPICM, CNRS, Ecole Polytechnique, Institut Polytechnique de Paris, 91120 Palaiseau, France}
\author{Vadim S. Nikolayev}\email[Contact email: ]{vadim.nikolayev@cea.fr}
\affiliation{SPEC, CEA, CNRS, Universit\'e Paris-Saclay, 91191 Gif-sur-Yvette Cedex, France}
\date{\today}
\begin{abstract}
The interfacial resistance to evaporation, in particular for the case of water, is a longstanding issue. The previous data on its main characterizing parameter, the accommodation coefficient, manifest a dispersion over three orders of magnitude. We study the evaporation of a thin liquid layer (microlayer) under the growing single bubble in saturated pool boiling of ultra-pure water at atmospheric pressure. However, the water contamination during the experiment cannot be excluded. The local and instantaneous data on interfacial thermal resistance are recovered from the synchronous local measurements of the microlayer thickness and the heater temperature. At each particular interfacial point, the resistance shows a linear increase in time suggesting the effect of accumulation of impurities at the interface. We compute the water mass evaporated at a given point of microlayer that should be proportional to the interfacial concentration of accumulated impurities. A clear increase of the interfacial thermal resistance with the evaporated mass is observed. This demonstrates a link of the temporal increase of the interfacial resistance (i.e. the evaporation deterioration) with the accumulation of impurities at the interface, which can explain the dispersion of the published data on the accommodation coefficient.
\end{abstract}

\maketitle
%\linenumbers
%\section{Introduction}\label{sec:intro}

In modern devices, liquid evaporation with subsequent condensation is widely used as the most efficient mean to transfer the energy and is called two-phase heat transfer. Its efficiency is due to the latent heat effect. The latent heat is absorbed at evaporation and then released at condensation in another place. Such phenomena occur e.g. in boiling, which is frequently used in many applications, from coffee making to energy conversion, phase separation in chemical engineering and cooling of power electronics. Probably the widest application is for cooling of electronic circuits by heat pipes that are now commonly used in every laptop computer and mobile phone. The latent heat is especially large for water so the devices that use this fluid are predominant. Modern technologies require progressively larger heat fluxes. The largest fluxes are achieved using thin liquid films due to property of pure two-phase fluids to maintain the temperature $T^i$ at the liquid-vapor film interface equal to that of saturation $T_{sat}$, which depends only on the vapor pressure. The temperature $T_w$ of the solid-liquid interface (that we will call simply wall) can be high, so the superheating $\Delta T=T_w-T_{sat}$ can be as large as $\sim \SI{50}{K}$. The transversal heat flux $q^i=k_l(T^i-T_{sat})/\delta$ (where $k_l$ is the liquid thermal conductivity and $\delta$ is the film thickness) can be huge because $\delta$ can be micrometric and even nanometric. Because of the film thinness, the heat flux is the same at the wall and at the liquid-vapor interface.

The heat flux can however be restricted by interfacial phenomena. To describe such a restriction, one commonly uses the \citet{Schrage} theory based on the fact that the vapor speed is limited by the thermal speed of gases. After linearization \cite{SWEP22},  this theory leads to an expression 
\begin{equation} \label{eq:Ti}
	T^i=T_{sat}+q^i R^i,
\end{equation}
where the interfacial thermal resistance is introduced in the form
\begin{equation} \label{eq:Ri}
	R^i=\dfrac{2-\mathcal{F}}{2\mathcal{F}} \dfrac{T_{sat}\sqrt{2\pi \mathcal{R}_g T_{sat}}}{\mathcal{L}^2\rho_v}
\end{equation}
under the assumption that the liquid density is much larger than that of vapor. Here, $\mathcal{L}$, $\mathcal{R}_g$ and $\rho_v$ denote the latent heat, specific gas constant and the vapor density, respectively. The accommodation coefficient  $\mathcal{F}$ describes the probability to cross the liquid-vapor interface for a molecule. The interfacial heat flux becomes  
\begin{equation}\label{eq:qi}
q^i = \frac{\Delta T}{\delta/k_l+R^i}
\end{equation}
and is limited now by the $R^i$ value. While the ``theoretical'' value $R^i_{theo}\equiv R^i(\mathcal{F}=1)$ is very small (e.g. for water at \SI{1}{bar} and $100^\circ$C it is \SI{0.064}{\micro\kelvin.\meter^2/\watt}), the accommodation coefficient different from unity leads to its increase. As a parameter (in many cases, a fitting parameter), the interfacial thermal resistance is used in every model of two-phase heat transfer that concerns high heat fluxes. Its value, often characterized in terms of $\mathcal{F}$, is a long-standing problem. While for most fluids $\mathcal{F}$  can be considered as unity or close to it, the water is an exception. \citet{Marek01}, and \citet{Davis06} reviewed numerous works and found that the measured $\mathcal{F}$ values vary between $10^{-3}$ and 1 for water. More recently, this range is reduced but the tendency is similar. Some authors \cite{Chandra20,Kazemi23} find $\mathcal{F}=1$  or the values of the order 0.1 \cite{Lu19} while most evaluations \cite{Giustini16,Bures22} result in the order $10^{-2}$. Some of us \cite{IJHMT24} found recently that $\mathcal{F}$ decreased in time monotonously by a factor of ten during evaporation until attaining a value of $\sim 10^{-2}$  (the interfacial resistance increased in the same proportion).

\begin{figure*}[htb]
	\centering
	\includegraphics[width=0.8\textwidth,clip]{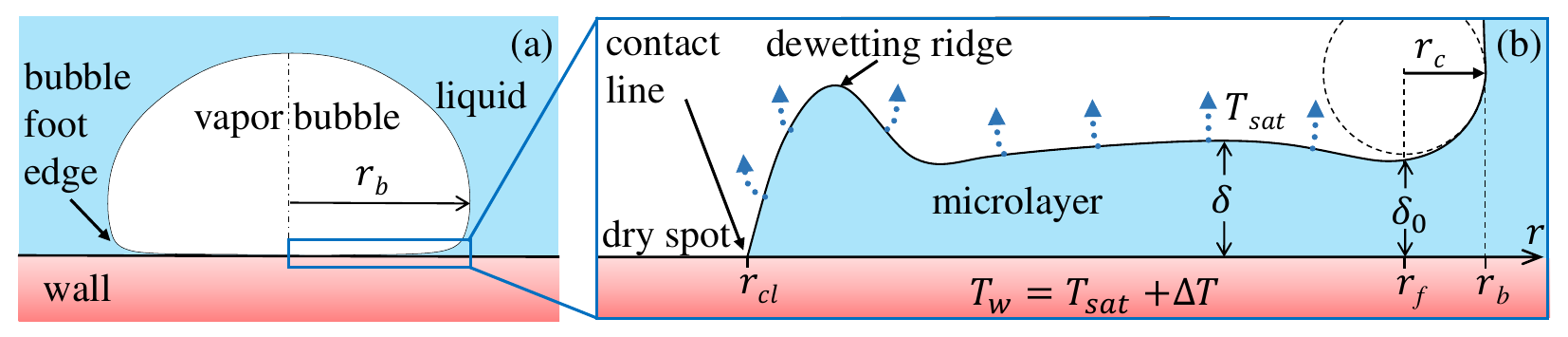}
	\caption{Schematics of a single bubble growing on a heated wall. (a) Bubble macroscopic view. (b) Microlayer.}
	\label{fig:nearwall}
\end{figure*}
Several reasons can lead to such a disparity. The first reason is likely a measurement error. The interfacial resistance can be measured only indirectly; careful experimental uncertainty analyses are however rare. Second, it can be an effect of the degree of water purity, specific to each particular work.

In heat transfer applications, the fluid can be rarely considered as pure. For instance, it is well-known that the fluid contamination by dissolved ``noncondensable'' gases (i.e. those that cannot change state from gas to liquid and vice-versa in the working temperature range), if not eliminated from the interface vicinity, substantially hinder the heat transfer \cite{Ghiaasiaan11,Ohashi20} at high heat fluxes even for very small concentrations. A boundary layer containing the noncondensable gas forms at the vapor interface side \cite{Lu19}. The vapor should diffuse through it to be evacuated or supplied from/to the interface in case of evaporation or condensation, respectively. Another sort of contaminants are hydrocarbons that are dispersed in the air \cite{Illing14} and are also known to deteriorate the boiling efficiency \cite{Song20}. This is a general issue that applies to most fluids. 

Another issue related to impurities is specific to water that possesses unique electro-chemical properties. In particular, its liquid-vapor interface is electrically charged, which can cause a sticking of the impurities to it and thus their accumulation \cite{Bjoerneholm16,Poli20}. In the presence of evaporation, the fluid molecules cross the interface while the impurities remain on it. Therefore, the impurities previously distributed in the liquid volume are eventually accumulated at the interface, leading to a high interfacial concentration. This phenomenon is expected to lead to the evaporation deterioration even for pure water as even a monolayer of impurity molecules can considerably decrease the interfacial phase change \cite{Hickman71,Barnes86,Miles16,Hartfield24}. According to \cite{Hickman71}, an addition of hydrocarbons (e.g. the behenic acid) in a proportion as small as one molecule per $4600\textup{~\AA}^2$ of the water interface caused an increase of liquid superheating near the interface (significant of the evaporation rate decrease) of 9\%.

\begin{figure*}[ht!]
\centering
\includegraphics[width=0.7\textwidth,clip]{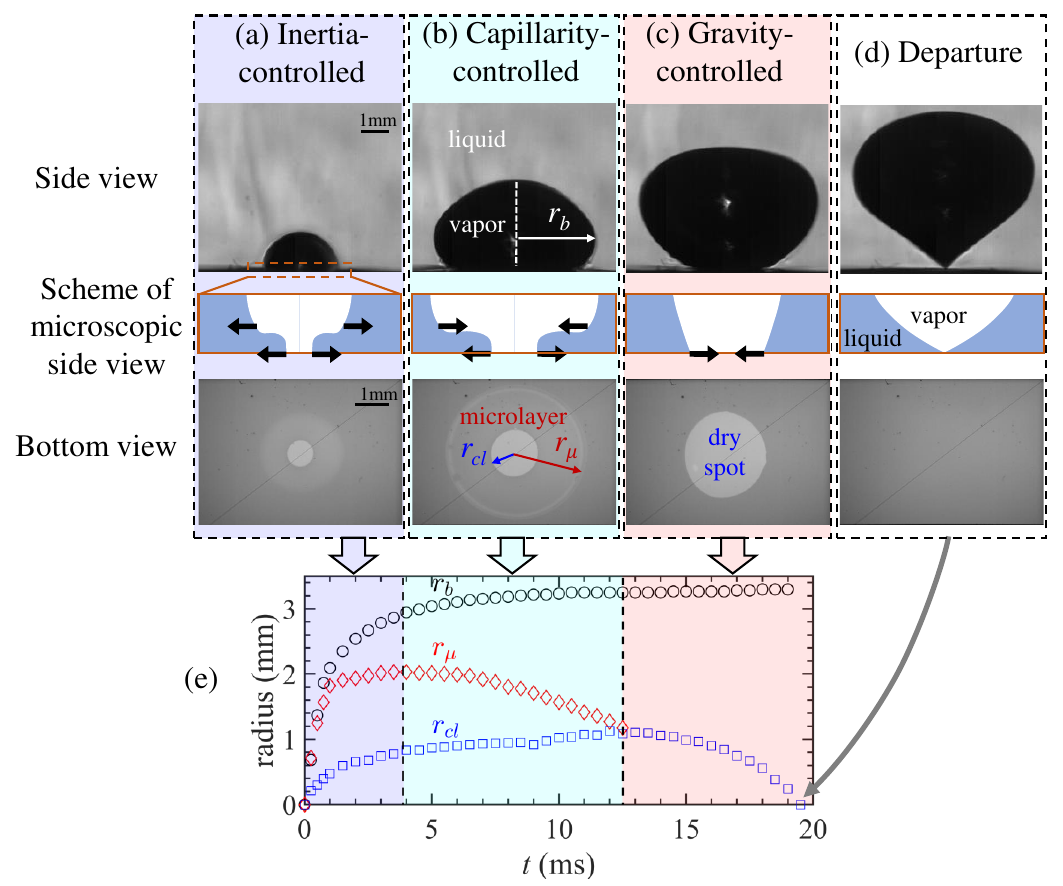}
	\caption{Bubble and microlayer dynamics for the case A. The top row shows the shadowgraphy bubble images from which one can extract the dynamics of the bubble radius $r_b$. The second (from the top)  row shows the corresponding near-wall microscopic geometry of the interface. The third row shows the corresponding bottom view images filmed through the transparent heater showing the dry spot radius $r_{cl}$ (i.e. that of the triple contact line) and the microlayer edge radius $r_\mu$. The bottom graph (e) shows the time evolution of $r_b,\, r_{cl}$ and $r_\mu$ from the bubble nucleation to departure.} 
	\label{fig:FigBubble}
\end{figure*}
In this paper we characterize the interfacial resistance by studying the microlayer evaporation at nucleate boiling. %Despite over a century of research, we still lack a fundamental understanding of it, primarily due to its multi-scale nature. Indeed, the micro- and nano-scale phenomena strongly affect the macroscopic boiling parameters  \cite{Kim09,SWEP22}. 
After the nucleation of a bubble on a heating wall, its fast growth can lead to the formation of a few \si{\micro\meter}-thick liquid layer (\autoref{fig:nearwall}) between the wall and the liquid-vapor interface of the bubble known as microlayer \cite{Cooper69}. The existence of microlayer was suggested in the pioneering work of \citet{Moore61} where micrometric homemade thermocouples measured high wall temperature variation that could be explained neither by liquid advection nor the contact line effect. \citet{Torikai67} performed a direct observation through the transparent heater: an electro-conductive glass. They noticed a dry area smaller than the bubble itself, which confirmed the existence of such a microlayer. \citet{Jawurek69} was the first to measure its thickness with laser interferometry. The microlayer plays a key role in boiling as it serves as a heat transfer bridge between the heated wall and the fluid, resulting in elevated evaporation rates that generate high heat fluxes, cool down the wall, and notably contribute to the overall bubble growth \cite{Myers05,Jung14,IJHMT24}. 

The bubble evolution from nucleation to departure is illustrated in \autoref{fig:FigBubble}. Around the nucleation site, a growing dry spot appears, creating a triple solid-liquid-vapor contact line on the wall. Its receding leads to the dry spot spreading, a key mechanism that triggers the boiling crisis \cite{PRL06,Zhang23b}. First, the bubble growth is extremely fast and for this reason is controlled by the liquid inertia. The inertial forces lead to the hemispherical bubble shape (\autoref{fig:FigBubble}a). If the bubble foot edge (cf. \autoref{fig:nearwall}a) recedes faster than the contact line, the microlayer forms. This is analogous to the dynamic wetting transition observed in capillaries \cite{Gao19,PRF23} where the contact line cannot follow a receding liquid meniscus.
As the bubble growth slows down, the inertial forces weaken and the capillary force (and the gravity to some extent) cause the reduction of microlayer length, i.e. the advancing of the bubble foot edge while the contact line keeps receding (\autoref{fig:FigBubble}b). At $r_\mu=r_{cl}$, the microlayer area becomes zero. After that, the gravity becomes the dominant force controlling the bubble dynamics (\autoref{fig:FigBubble}c); the contact line advances (i.e. moves back to the nucleation site), which eventually leads to the bubble departure (\autoref{fig:FigBubble}d). 
%Unlike the microlayer area, the low thermal conductivity of vapor cuts the wall heat flux in the dry spot area. Therefore, the fundamental understanding of microlayer and contact line dynamics is essential for predicting the overall boiling heat transfer. 
%Nucleate boiling took place on ITO and heating was ensured by continuous infra-red (IR) laser heating. 

In a recent work, \citet{JFM24Boiling} showed that in spite of the bubble dynamics governed by inertia during the microlayer formation, the latter is controlled by the interplay of viscous and surface tension forces and is analogous to the deposition of a liquid film by the receding meniscus (Landau-Levich film) on a flat plate \cite{LL42} or in a capillary \cite{Bretherton}. The thickness of such a film is defined by the capillary number $Ca_b=\mu U_b/\sigma$ where $U_b=\dot r_b$, where dot means the time derivative, represents the speed of the receding bubble foot edge (cf. \autoref{fig:nearwall}) with the curvature $r_c^{-1}$ :
\begin{equation} \label{eq:delta0}
	\delta_0=1.34r_cCa_b^{2/3}.
\end{equation}
Here, $\mu$ and $\sigma$ stand for the dynamic viscosity and the surface tension, respectively. $\delta_0=\delta_0(r)$ denotes the initial microlayer thickness, i.e. the thickness right after microlayer formation at a point defined by its radial distance $r$ from the bubble center. This thickness is observed near the bubble foot edge (\autoref{fig:nearwall}b). Note that Eq.~\eqref{eq:delta0} is valid for $Ca_b\ll 1$ and that $r_c$ is much smaller than the bubble radius $r_b$ shown in \autoref{fig:nearwall}a because the bubble foot is flattened by the inertial forces (\autoref{fig:FigBubble}a). %Therefore, a portion of bubble interface between the foot and the dome has a larger curvature $r_c^{-1}$ than the bubble dome, as shown in \autoref{fig:nearwall}. %To compute $\delta_0$, the values of $r_b$ and $U_b$ are obtained from the experimental measurements (see \autoref{sec:initial-exp}). 
%In Eq.~\eqref{eq:delta0}, $\delta_0$ can vary with the distance $r$ from the bubble center. 

\section{Results and discussion}%\label{sec:heatflux}

To understand the origin of a large interfacial thermal resistance, one needs first to introduce and evaluate the initial microlayer thickness.

\subsection{Initial microlayer thickness}\label{sec:deltao}

The microlayer evaporation is studied here by using an experimental setup \cite{JFM24Boiling,IJHMT24,EuroTherm24Exp} coupling simultaneous measurements of (i) the microlayer thickness $\delta(r,t)$ with white light interferometry (WLI), (ii) the wall temperature $T_w$ with infra-red thermography (IRT), and (iii) the macroscopic bubble shape with shadowgraphy. There is almost no flow in the nearly flat part of microlayer (whose thickness is measurable by WLI) because of the strong viscosity effect \cite{JFM24Boiling}, which means that the evolution of $\delta$ is governed by the local evaporation mass flux $j$ 
\begin{equation} \label{eq:balance}
	\dot\delta = -j/\rho_l
\end{equation}
where $\rho_l$ is the liquid density. On the other hand, the interfacial energy balance can be expressed as  
\begin{equation}\label{mfl}
\mathcal{L}j = q^i,
\end{equation}
where $\mathcal{L}$ stands for the latent heat of vaporization. Combined together, Eqs. (\ref{eq:qi}, \ref{eq:balance}, \ref{mfl}) yield 
\begin{equation}\label{eq:govern}
\dot\delta = -\frac{1}{\rho_l\mathcal{L}}\frac{\Delta T}{\delta/k_l +R^i},
\end{equation}
which shows that the microlayer evaporation is entirely local phenomenon like melting of a solid under the influence of a local heating. 

\begin{figure}[tb]
	\centering
\includegraphics[width=0.8\columnwidth,clip]{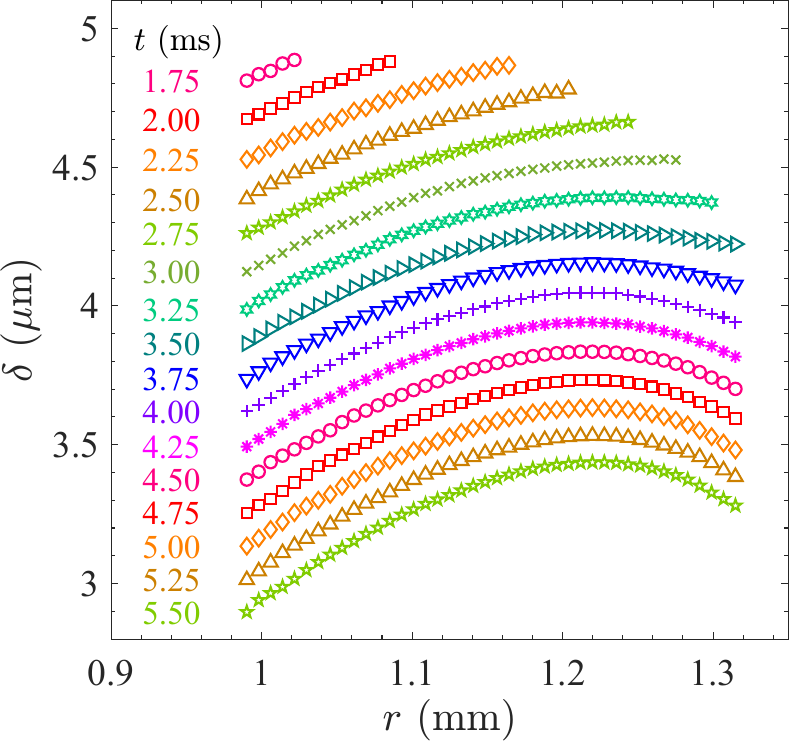}
	\caption{Evolution of microlayer profile over time for the case B with $q_a''=\SI{80.7}{\kilo\watt/\meter^2}$.} 
	\label{fig:delta28}
\end{figure}
Figure \ref{fig:delta28} depicts an example of the spatio-temporal evolution of the microlayer thickness $\delta$ measured with WLI. One can observe that the radial extent of microlayer ($r_{cl},r_\mu$) grows at early times, $t \leq\SI{3.75}{\milli\second}$. This growth reflects the process of microlayer formation illustrated in \autoref{fig:FigBubble}a, where the microlayer edge radius $r_\mu$ increases faster than $r_{cl}$ (\autoref{fig:FigBubble}e). Recent numerical simulations \cite{JFM24Boiling,Giustini24,Guion18,Urbano18} show that the microlayer consists of two parts: a dewetting ridge near the contact line followed by a nearly flat part, as illustrated in \autoref{fig:nearwall}b. All the results on microlayer thickness presented below correspond to this flatter part because the much steeper ridge cannot be resolved by WLI due to the maximum slope limitation typical for the interferometry measurements \cite{JFM24Boiling,Choi24}. The limitation is clearly seen in \autoref{fig:delta28}. 

One can see that the microlayer thins over time. The microlayer profile is shifted vertically due to evaporation, almost without any other modification, which confirms the absence of flow along it in agreement with the numerical simulations \cite{Hansch16,Urbano18}. We are however interested to know the initial microlayer thickness that enters Eq.~\eqref{eq:delta0}. For this, one needs to return from the instant $t_m$ where its thickness $\delta_m(r)=\delta(t=t_m,r)$ has been measured to the time of microlayer formation $t_f=t_f(r)$ where its thickness was $\delta_0$:
\begin{equation}\label{delta_dep}
\delta_0=\delta(t=t_f).
\end{equation}
\begin{figure*}[ht!]
	\centering
\subfloat[$\delta_0$ for different cases. The case B is plotted for $q_a''=\SI{80.7}{\kilo\watt/\meter^2}$. For each case, the width of shaded rectangle indicates the extent of microlayer ($r_{cl},r_\mu$) when $r_\mu$ is at its maximum.]{\includegraphics[width=0.8\columnwidth,clip]{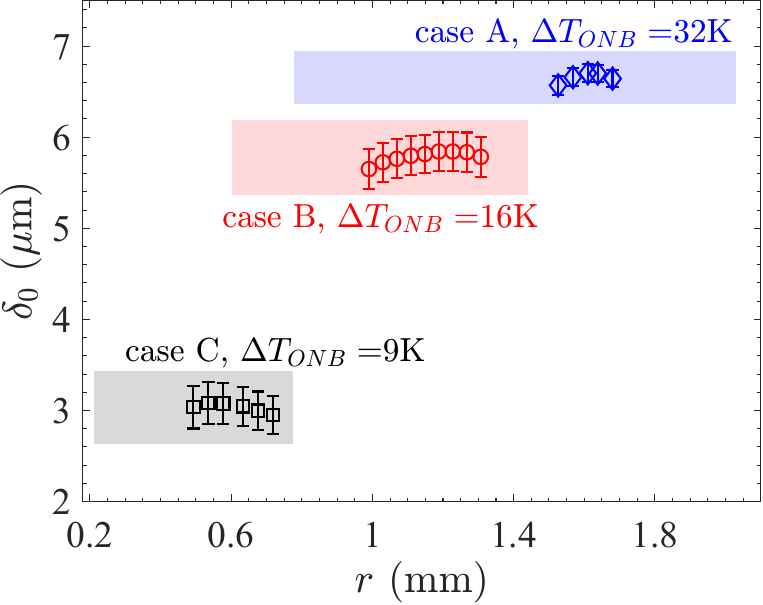}\label{fig:deltaoDT}}\hspace{5mm}%
\subfloat[Effect of $q_a''$ on $\delta_0$ for the case B.]{\includegraphics[width=0.8\columnwidth,clip]{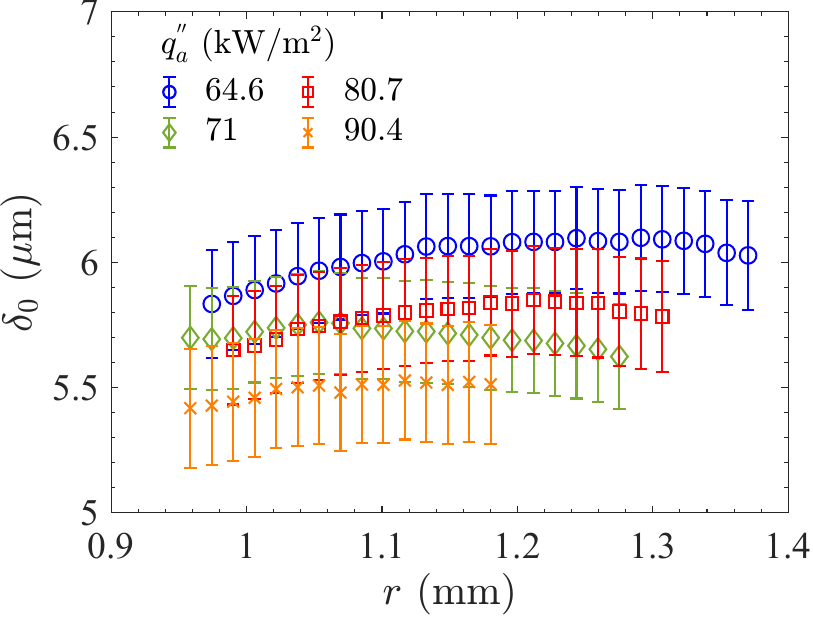}\label{fig:deltao_heating}}
	\caption{$\delta_0$ as a function of the radial position $r$.} 
	\label{fig:deltao}
\end{figure*}
To obtain a good accuracy, it is advantageous to choose $t_m$ as close to $t_f$ as possible, i.e. as a time instant of the first $\delta$ measurement at a given point $r$. Consider, for instance, $r=\SI{1.1}{\milli\meter}$ in \autoref{fig:delta28} where the microlayer is formed between 2 and \SI{2.25}{\milli\second}; we choose $t_m(r=\SI{1.1}{\milli\meter})=\SI{2.25}{\milli\second}$. 

It is convenient to define the inverse to $t_f(r)$ function defining the point $r=r_f(t)$ where the thickness given by Eq.~\eqref{eq:delta0} is formed at the time $t$, which allows us to express $\delta_0$ as a function of $r$.  During the formation of the microlayer,  its thickness $\delta_0$ is achieved somewhere behind the bubble foot edge, so $r_f< r_b$, cf. \autoref{fig:nearwall}b. According to the numerical analysis of the film deposition by a meniscus of variable speed \cite{JFM21} (cf. \autoref{fig:nearwall}b for the geometrical rationale), 
\begin{equation}\label{eq:rdep}
	r_f\simeq r_b-r_c.
\end{equation}

For $\delta_0$ reconstruction, one can use Eq.~\eqref{eq:govern}. However, the $R^i$ value entering it is \emph{a priori} unknown. We know only that $R^i\ll \delta/k_l$ right after the microlayer formation \cite{IJHMT24} so it can be neglected within the time interval $(t_f, t_m)$. This assumption is applied only for evaluation of $\delta_0$.  Inside such a small time interval, one can replace $\Delta T$ by its time average $\overline{\Delta T}$. The analytical solution of Eq.~\eqref{eq:govern} yields
\begin{multline}\label{eq:delta0Exp}
	\delta_0(r)=\Bigg\{\delta[r,t_m(r)]^2\\+2\frac{k_l\overline{\Delta T}(r)  (t_m(r)-t_f(r))}{\mathcal{L}\rho_l}\Bigg\}^{1/2}.
\end{multline}
One can solve now the set of three equations (\ref{eq:delta0}, \ref{eq:rdep}, \ref{eq:delta0Exp}) to obtain $\delta_0$, $t_f$ and $r_c$ for each point $r$ (for details, cf. Supplementary information, sec.~II).

Figure~\ref{fig:deltaoDT} shows $\delta_0$ for the present (B) case and compares it to the data published earlier (cases A \cite{IJHMT24} and C \cite{EuroTherm24Exp}), corresponding to different values (\autoref{tab:testcases}) of the Onset of Nucleate Boiling superheating $\Delta T_{ONB}$ required to nucleate a bubble. The bubble growth rate on a massive heater is controlled mainly by $\Delta T_{ONB}$ because of the heater thermal inertia; the heat flux applied by the heater (we mention its reference value $q_a''$ defined in Supplementary information, sec.~I) has only a weak effect on the bubble growth rate \cite{IJHMT24}. The weak effect of $q_a''$ on $\delta_0$ is demonstrated in \autoref{fig:deltao_heating}. 

Three important results are apparent from \autoref{fig:deltao}. First, a higher $\Delta T_{ONB}$ causes a thicker microlayer. Second, the microlayer shows a ``bumpy'' profile. Third, both the microlayer extent and the position of its maximum thickness increase with $\Delta T_{ONB}$. The first feature can be explained with the Landau-Levich theory of microlayer formation if one mentions that a faster bubble growth observed at a higher $\Delta T_{ONB}$ causes a higher capillary number $Ca_b$ resulting in a higher $\delta_0$ according to Eq.~\eqref{eq:delta0}. The microlayer bump was also observed by other groups \cite{Chen17,Jung18}.  The maximum thickness is predicted by Eq.~\eqref{eq:delta0} because $r_c$ grows similarly to $r_b$ while $Ca_b\propto \dot r_b$ decreases over time (cf. \autoref{fig:FigBubble}e) \cite{JFM24Boiling}. The third feature is explained, on one hand, by a larger bubble size (i.e., departure radius) for a larger superheating \cite{ATE25}, and on the other, by a stronger bubble flattening by inertial forces at a faster bubble growth.

\subsection{Interfacial thermal resistance}\label{sec:Ri}

\begin{figure*}[ht!]
	\centering
\subfloat[The ratio $R^i/R^i_{theo}$ as a function of the microlayer evaporation time.]{\includegraphics[width=0.75\columnwidth,clip]{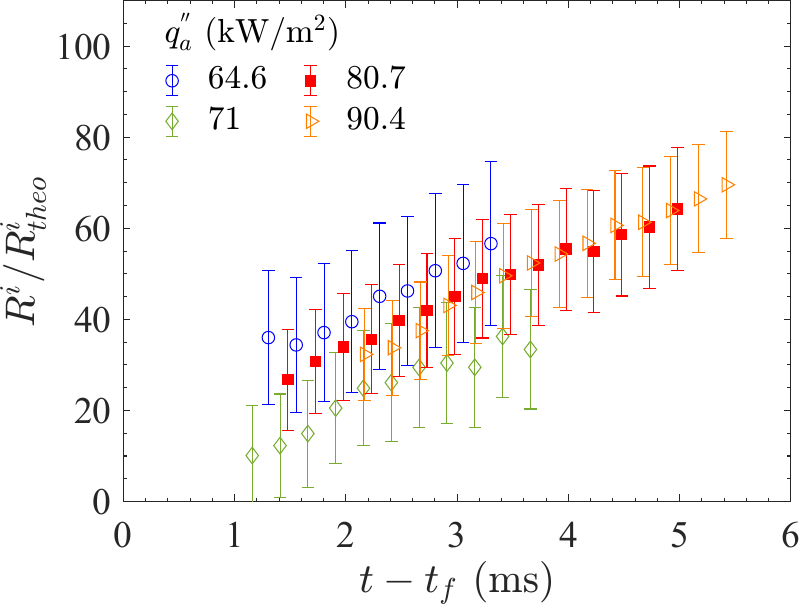}\label{fig:Rius}}\hspace{7mm}%
\subfloat[$\mathcal{F}$ as a function of the microlayer evaporation time. The data from previous works \cite{Giustini16,Lu19,Bures22,IJHMT24} are shown with lines.]{\includegraphics[width=0.75\columnwidth,clip]{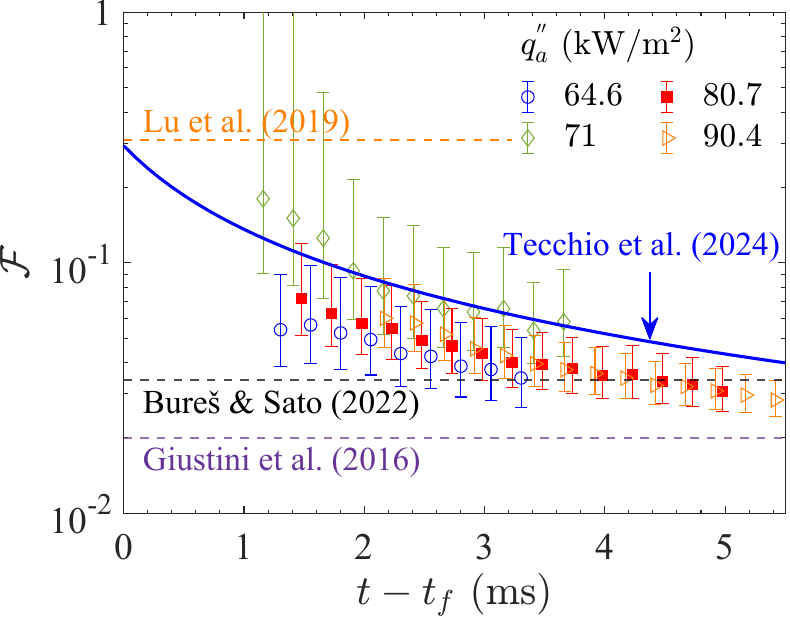}\label{fig:Fus}}
	\caption{The evolution of the interfacial thermal resistance and the corresponding accommodation coefficient for the case B at $r=\SI{1.03}{\milli\meter}$.} 
	\label{fig:RandF_us}
\end{figure*}
The study of the microlayer gives a rare opportunity for the precise interfacial resistance evaluation. The method used here to obtain $R^i$ is different from a more classical approach  \cite{Bures22} in which $R^i$ is determined by reconstructing from the $T_w$ data the heat flux $q^i$ and using it in Eq.~\eqref{eq:qi}. This requires solving a mathematically ill-posed heat transfer problem \cite{Cattani23} inside the porthole, which results in big $q^i$ fluctuations \cite{IJHMT24} if not properly regularized. To reduce the uncertainty, we compute here $R^i$ directly from Eq.~\eqref{eq:govern}, where the $\delta$ and $\Delta T$ values come from the WLI and IRT measurements, respectively. We also reprocess our previously obtained data for the cases A and C. The $\dot\delta$ term is calculated by using the $\delta$ values for two subsequent times. The determined values of $R^i$ measurement are thus both local and instantaneous. No phenomena hindering the determination of interfacial resistance exist here, unlike other experimental approaches \cite{Lu19} where e.g. heat losses had to be carefully evaluated. 

Figure~\ref{fig:Rius} shows the time evolution of $R^i/R^i_{theo}$ at a fixed location $r$ evaluated from the microlayer deposition at $t=t_f(r)$; the latter time is computed above together with $\delta_0$. One observes the interfacial thermal resistance increase at microlayer evaporation independently on $q_a''$. The corresponding data on the accommodation coefficient $\mathcal{F}=2/(R^i/R^i_{theo}+1)$ is shown in \autoref{fig:Fus} together with some previous data for comparison. One observes that $\mathcal{F}$ monotonously decreases in time. These results support the idea \cite{IJHMT24} about the $R^i$ increase due to the accumulation of impurities at the liquid-vapor interface during evaporation. To verify this hypothesis, one needs to compute the amount of accumulated impurities at a given interfacial point. It would be however difficult to measure this quantity directly. One can assume that the impurity spatial distribution is initially homogeneous inside the microlayer; while it evaporates, the impurities that were previously in the liquid stick to the interface, possibly because it is charged as discussed above. While the water molecules cross the interface, the impurities accumulate there. Therefore, the concentration of impurities at a given point $r$ of interface and at the time $t$  should be proportional to the total water mass per unit area $\mathcal{M}$ evaporated at the point $r$ from the microlayer formation until the time $t$,
\begin{equation}\label{Mfl}
 \mathcal{M}=\int_{t_f(r)}^{t}j(r,t)dt.
\end{equation}
By performing the integration with the account of Eqs.~(\ref{eq:balance}, \ref{delta_dep}), one finally obtains 
\begin{equation}\label{Mexp}
  \mathcal{M}(r,t)=\rho_l[\delta_0(r)-\delta(r,t)].
\end{equation}
This formula justifies the necessity of the above computation of the initial microlayer thickness. 
\begin{figure}[ht!]
	\centering\includegraphics[width=0.9\columnwidth,clip]{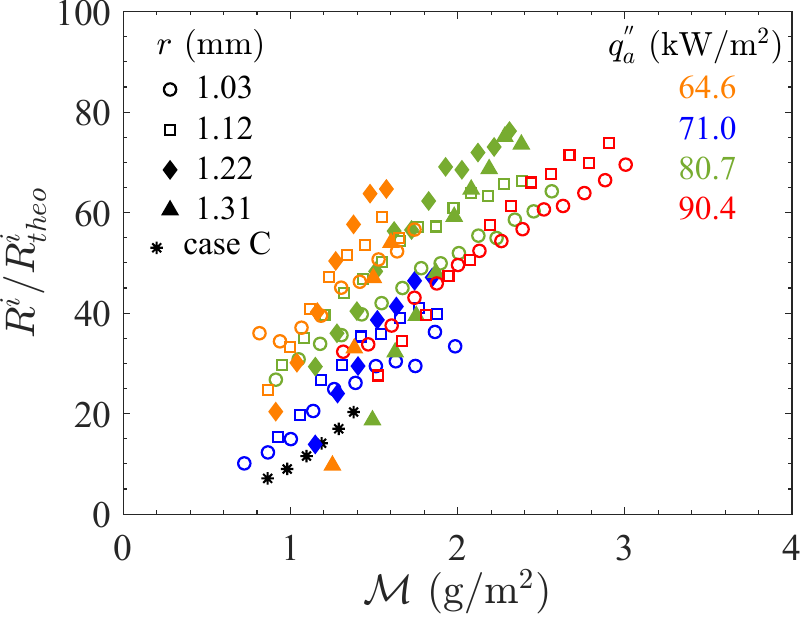} 
	\caption{Correlation of the interfacial thermal resistance with the integral evaporated water mass for the case B, for different locations, times and heat fluxes and for the case C at $r=\SI{0.67}{\milli\meter}$.} \label{fig:RiMass}
\end{figure}

The correlation of $R^i(r,t)$ with $ \mathcal{M}(r,t)$ is given in \autoref{fig:RiMass}  for different heat fluxes $q_a''$, positions $r$ and times (similar characters for increasing $\mathcal{M}$ correspond to increasing $t$). One can see that the points always follow an increasing $R^i(\mathcal{M})$ trend. This shows the proportionality of $R^i$ to the local instantaneous interfacial concentration of impurities, which constitutes the central result of this paper.
\begin{figure*}[htb]
	\centering
	\includegraphics[width=0.8\textwidth,clip]{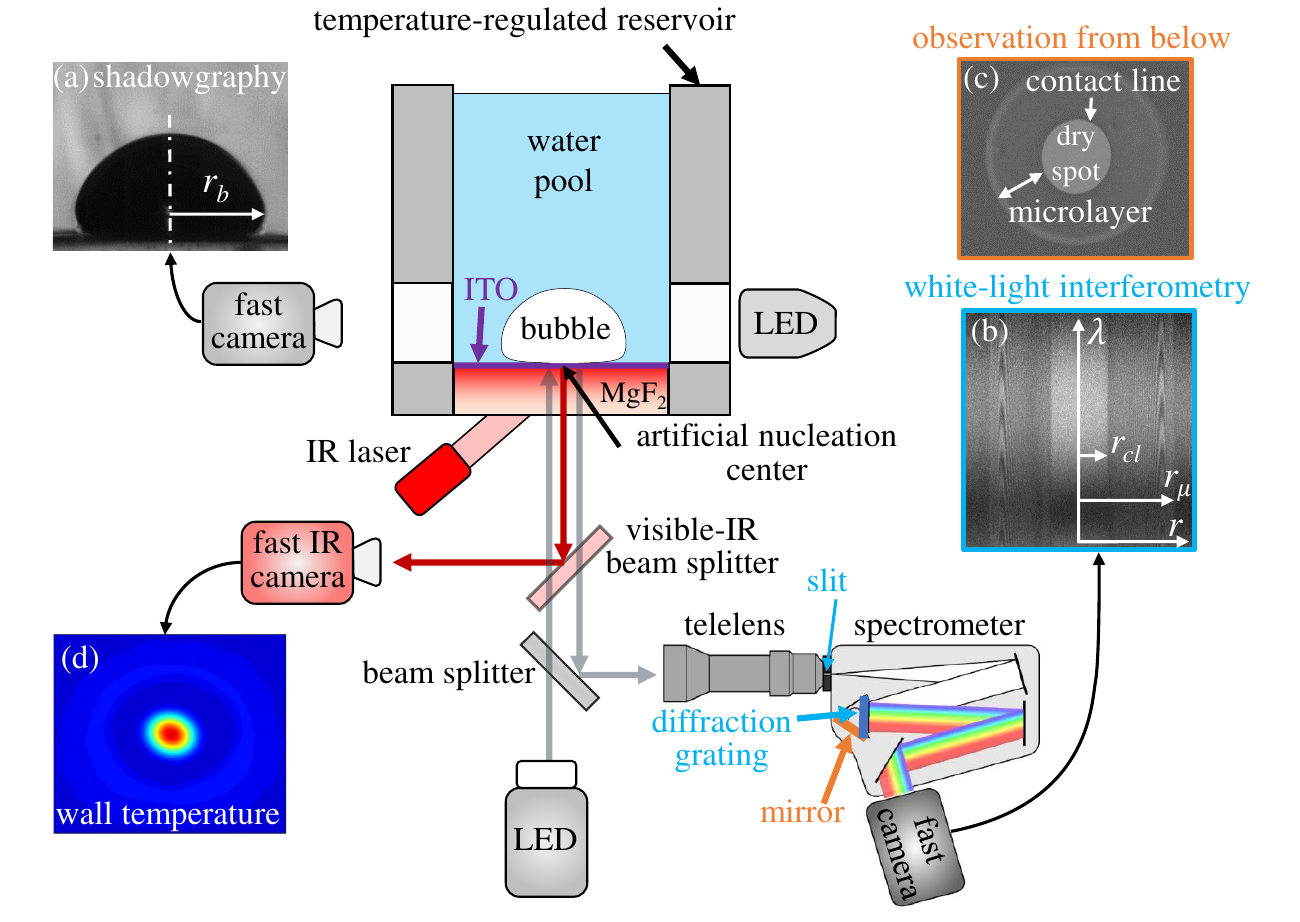}
	\caption{Schematics of the experimental setup and experimentally obtained images.} \label{fig:experiment}
\end{figure*}
There are three main reasons for the dispersion of the data points in \autoref{fig:RiMass}. The first reason is the experimental error (see Supplementary information, sec.~III). The second reason concerns a possible heterogeneity of impurity concentration. It can be formed in the liquid volume after the previous bubble departure. The third reason is a possible slight variation of water purity in different experiment sessions.

\section{Methods}\label{sec:initial-exp}

\begin{table*}[htp]
	\centering
	\caption{The list of experimental cases. The lengths $h$ and $d$ stand for the cavity depth and its mouth diameter, respectively; $D$ is the laser hot-spot diameter.}
	\begin{ruledtabular}\begin{tabular}{ccccccccc}
		
		case & nucleation center & $d$ ($\si{\micro\meter}$) & $h$ ($\si{\micro\meter}$) & $\beta$ (deg.) & $\Delta T_{ONB}$ ($\si{\kelvin}$) & $q_a''$ ($\si{\kilo\watt/\meter^2}$) & $D$ (\si{\milli\meter})\\
		A \cite{IJHMT24} & natural defect & - & - & - & 31.5 & 1100 & 1.5\\
		B & cavity & 38.6 & 70.5 & 6.7 & 15.4 & 64.6-113.0 & 6.3\\ 
		C \cite{EuroTherm24Exp} & cavity & 26.7 & 18.3 & 4.6 & 9.2 & 200 & 2.5\\ 
		
	\end{tabular}\end{ruledtabular}\label{tab:testcases}
\end{table*}

The experimental installation, measurement techniques and their calibration are described elsewhere \cite{JFM24Boiling,IJHMT24}. Only a concise description is given here.  The set-up schematics is presented in \autoref{fig:experiment}. We used the ultra-pure water purified with the Milli-Q IQ 7003 apparatus. Its electrical resistivity is \SI{18.2}{\mega\ohm.\centi\meter} and it contains only \SI{2}{ppb} of total organic carbon. The boiling cell consists of a water pool surrounded by a jacket regulating the water temperature to the saturation value ($100^\circ$C). The water pool is open to the atmosphere (which can possibly lead to water contamination) but is boiled for several hours for degassing before the experiments.

A \SI{950}{\nano\meter}-thick indium-tin oxide (ITO) film deposited on the \SI{3}{\milli\meter} magnesium fluoride (MgF$_2$) optical porthole serves as a boiling surface. The ITO film is deposited by using a radio-frequency magnetron sputtering system. MgF$_2$ is transparent to both visible and infrared (IR) light, whereas ITO is transparent to visible light but opaque in the IR spectrum that it absorbs. The growth of a single bubble at a time is obtained by heating up the ITO film locally with an IR (\SI{1.2}{\micro\meter}) continuous laser. The laser beam has a Gaussian-like profile. The wall heat flux $q_a''$ can be varied among different experimental cases (\autoref{tab:testcases}) by changing the beam diameter and the laser power. 

In this study, we compare the data obtained with three different boiling surfaces. \autoref{tab:testcases} summarizes their parameters. In the case A \cite{JFM24Boiling,IJHMT24}, the bubble nucleates on a natural point defect located by scanning the otherwise nanometrically smooth ITO surface with the IR laser beam. On the other two surfaces (B and C), the bubble nucleates on an artificial cavity fabricated on the MgF$_2$ porthole by the focused ion beam (FIB) technology prior to the ITO deposition. The cavity shape is weakly conical (almost cylindrical) \cite{EuroTherm24Exp}; the angle $\beta$ gives a difference of the lateral cavity walls from vertical. The use of different boiling surfaces and cavity dimensions allows us to obtain a wide range of wall superheating  values $\Delta T_{ONB}$ at bubble nucleation.

The boiling cycles follow a repetitive pattern featuring bubble nucleation, growth period, detachment and waiting period \cite{IJHMT24,EuroTherm24Exp}. The repeatability is very good for all the cases. 

The bubble macroscopic shape and its radius $r_b(t)$ (\autoref{fig:experiment}a, \autoref{fig:FigBubble}e) are obtained by sidewise shadowgraphy with a spatial resolution of \SI{32}{\micro\meter/px} for the case A and \SI{9}{\micro\meter/px} for the cases B and C. The microlayer thickness is measured by WLI. A collimated white light is shone perpendicularly to the MgF$_2$ porthole from below. The reflected light is directed with a beam splitter towards a spectrometer. At the entrance of the spectrometer, a slit defines a scanning line. The setup is aligned in such a way that the scanning line passes through the nucleation center. The light is dispersed by a diffraction grating inside the spectrometer; its camera records the spectral fringe map (\autoref{fig:experiment}b). It results from the interference of the rays reflected from interfaces between MgF$_2$/ITO, ITO/micro\-layer and microlayer/vapor. In \autoref{fig:experiment}b, the ordinate corresponds to the wavelength $\lambda$, and the abscissa to the physical coordinate on the wall along the scanning line. The diffraction grating can be replaced by a mirror to observe the bubble from below (\autoref{fig:experiment}c). Thanks to the low roughness of the ITO surface, in all the cases the bubbles show a cylindrical symmetry (\autoref{fig:experiment}c) which allows us to associate the distance along the scanning line to the radial coordinate $r$. The microlayer thickness profile $\delta(r,t)$ is determined by comparing the fringe map (\autoref{fig:experiment}b) with the two-beam interference model \cite{Tecchio22}. From \autoref{fig:experiment}b, one can also obtain the dry spot (contact line) radius $r_{cl}$ (\autoref{fig:FigBubble}e) as a radius where the optical contrast changes sharply due to different reflectivity from the ITO interface in the areas of contact with vapor and liquid. From the optical analysis, one can define the microlayer radius $r_{\mu}$ (\autoref{fig:FigBubble}e) as a maximal radial coordinate, within which the interference fringes (giving a brighter on average image than the bulk liquid) appear. The wavelength bandwidth in WLI is $\SI{444.6}{\nano\meter}\leq\lambda\leq\SI{587.7}{\nano\meter}$, and the spectral resolution is \SI{0.21}{\nano\meter/px}. The WLI spatial resolution along $r$ is \SI{14}{\micro\meter/px} for the case A and \SI{7.9}{\micro\meter/px} for the cases B, C. The resolution fixes the maximum slope of the vapor-liquid interface that can be measured \cite{JFM24Boiling}. The maximal slopes at $\lambda=\SI{500}{\nano\meter}$  are $0.38^\circ$ for the case A and $0.76^\circ$ for the cases B, C. With such an improvement, it is possible to measure the profile of $\simeq 30$\% of the total microlayer length (\autoref{fig:deltaoDT}).

The radiation emitted by ITO in the range 3--\SI{5}{\micro\meter} is captured by the fast IR camera used for IRT. A resistance temperature detector measures the water pool temperature near the boiling surface and serves for IRT calibration. The spatial resolution of the IRT is \SI{84}{\micro\meter/px} for the cases A and C, and \SI{100}{\micro\meter/px} for the case B. 

All the acquired images are recorded synchronously at \SI{4000}{fps} rate. %Both the calibration and validation of WLI and IRT are described in \cite{Tecchio22} and the measurement uncertainties are provided in \autoref{tab:uncertainty}.

%% ----------------------------------------------------------------------------------------
%% Conclusion
\section{Conclusion}\label{sec:conclusion}

The study of microlayer in boiling offers a rare opportunity for the local and instantaneous evaluation of the interfacial thermal resistance and accommodation coefficient. Their measurement does not rely on estimations of any kind and is thus quite precise. %The ultra-pure water is used. 
The interfacial thermal resistance has been evaluated for a broad range of well controlled experimental conditions. In agreement with previously published results, the interfacial thermal resistance grows during the microlayer evaporation corresponding to a decrease of the accommodation coefficient. It is known however from the literature that the evaporation is extremely sensitive to the interface contamination. Even a monolayer of molecules can affect it. It is suggested that the growth of the interfacial thermal resistance is caused by accumulation of impurities at the vapor-liquid interface during evaporation. We argue that because of the specific electrochemical properties of water (its interface is charged),  the impurities stick to the interface (i.e. do not diffuse). Their interfacial concentration should thus be proportional to the locally evaporated integral water mass. The experimental data confirm this hypothesis showing that the interfacial thermal resistance (i.e. the barrier to evaporation) grows with the evaporated water mass. The temporal deterioration of the water evaporation efficiency can thus be caused by various interfacial contamination even for the ultra-pure water (that can however be contaminated during the experiment), which explains the disparity of the published data on the interfacial resistance of water.

\section*{Acknowledgments}

C.L.H. and B.Z. thank DM2S/CEA for their PhD and post-doctoral grants.

\section*{Author contributions}

P.B. and P.R.C. performed the ITO deposition and characterized the ITO properties.
C.L.H. and C.T. designed and conducted the experiments.
C.L.H, B.Z., and C.T. developed the post-processing techniques.
C.L.H, B.Z., C.T., and V.S.N. performed the experimental analysis.
V.S.N., C.T., and C.L.H. wrote the manuscript. All authors participated in discussing and reviewing the manuscript.

\section*{Competing interests}

The authors declare no competing interests.
\section*{Additional information}
Supplementary information for this paper is available

Correspondence and requests for materials should be addressed to C.T. or V.S.N.
%% ----------------------------------------------------------------------------------------
\bibliographystyle{elsarticle-num-names}
\bibliography{ContactTransf,Taylor_bubbles,PHP,Books}

\end{document}

\title{Deterioration of water evaporation by impurity accumulation at the liquid-vapor interface during nucleate boiling:\texorpdfstring{\\}{}SUPPLEMENTARY INFORMATION}
\author{Corentin Le Houedec}
\affiliation{STMF, Universit\'e Paris-Saclay, CEA, 91191 Gif-sur-Yvette Cedex, France}
\author{Cassiano Tecchio}
\affiliation{STMF, Universit\'e Paris-Saclay, CEA, 91191 Gif-sur-Yvette Cedex, France}
\author{Bo\v stjan Zajec}
\affiliation{STMF, Universit\'e Paris-Saclay, CEA, 91191 Gif-sur-Yvette Cedex, France}
\author{Pere Roca i Cabarrocas}
\affiliation{LPICM, CNRS, Ecole Polytechnique, Institut Polytechnique de Paris, 91120 Palaiseau, France}
\author{Pavel Bulkin}
\affiliation{LPICM, CNRS, Ecole Polytechnique, Institut Polytechnique de Paris, 91120 Palaiseau, France}
\author{Vadim S. Nikolayev}
\affiliation{SPEC, CEA, CNRS, Universit\'e Paris-Saclay, 91191 Gif-sur-Yvette Cedex, France}
\date{\today}

\maketitle
%\linenumbers

\section{Reference heat flux for the Gaussian laser profile}\label{sec:appA}
\begin{figure}[ht!]
	\centering
	\includegraphics[width=0.8\columnwidth,clip]{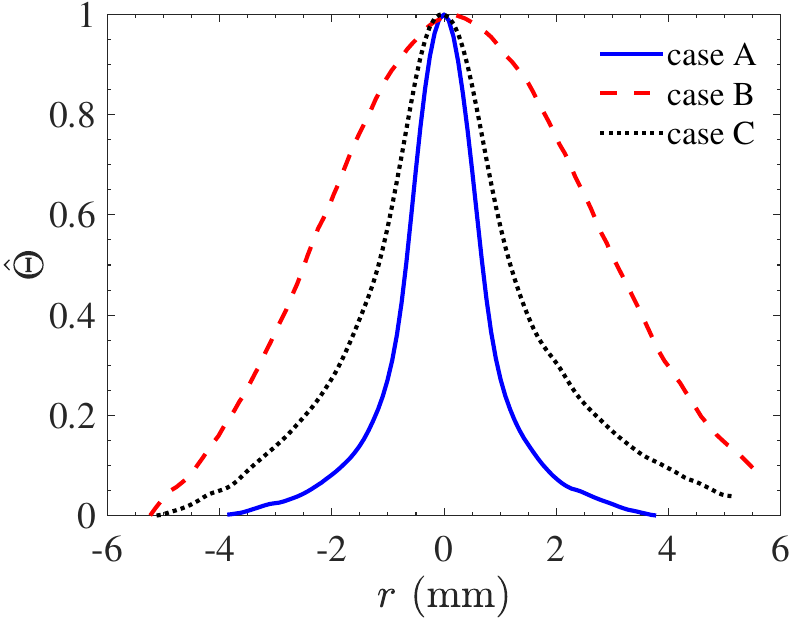}
	\caption{Dimensionless temperature profiles along the scanning line for different cases of Table I of the main article. For the case B, $q_a''=$ \SI{80.7}{\kilo\watt/\meter^2}.} 
	\label{fig:THETAprofiles_DT}
\end{figure}
The absorbed by ITO  IR laser Gaussian-like power profile can be characterized by two parameters: a half width size $D$ and the laser incident angle on the sample $\alpha$. Since it is different from $0^\circ$, the heating power profile is slightly elliptical. To determine $D$ and $\alpha$, we consider the 2D superheating temperature field $\Delta T(x,y)$ just before the nucleation event and define the dimensionless temperature $\hat{\Theta}=\Delta T(x,y)/\Delta T_{ONB}$. It is assumed that the laser is centered at the nucleation site location. Then we select the dimensionless isotherm $\hat{\Theta}=1/\sqrt 2$ and fit an ellipse to it. $D$ is determined as its minor diameter. %The ellipticity gives  $\alpha$. 
The heat flux $q_a''$ is determined as the laser power times the ITO absorption \cite{Tecchio22} divided by $\pi D^2/4$. Such a definition has been used by \citet{Mehrvand17,Tecchio22}, and gives an estimate that is sufficient in our case because the flux is used only for reference. Profiles of $\hat{\Theta}$ just before the nucleation along the scanning line are depicted in \autoref{fig:THETAprofiles_DT}. Note that the minor diameter lies at an angle to the scanning line \cite{Tecchio22}, so $D$ should not be evaluated with this graph.

\section{Data processing algorithm}\label{sec:appB}
\begin{figure}[htb]
	\centering\includegraphics[width=0.9\columnwidth,clip]{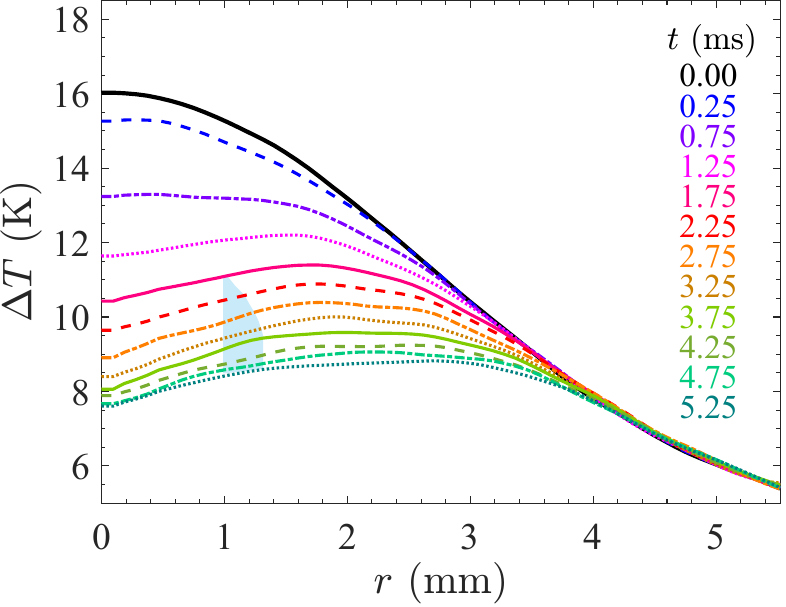} 
	\caption{Time evolution of $\Delta T$ profile for the case B with $q_a^{''}=\SI{80.7}{\kilo\watt/\meter^2}$. The blue shading indicates the extent of the observable part of microlayer (i.e. the fringes) for each time.}
	\label{fig:AppC:DT28}
\end{figure}
The $\Delta T$ spatial dependence has to be smoothed and linearly interpolated prior to the computation (\autoref{fig:AppC:DT28}) because WLI and IRT spatial resolutions differ. Similarly,  the fit of the $r_b(t)$ data is used in what follows. The set of equations (4,9,10) of the main article is solved for $\delta_0$, $t_f$ and $r_c$ at each point $r$ (that is in fact $r_f$) with the following algorithm:
\begin{enumerate}
	\item Initialize $r_c=0$.
	\item From Eq.~(9), calculate $r_b$.
	\item This $r_b$ value should correspond to the time $t_f$ of microlayer formation at the point $r$. We thus evaluate $t_f$ by inversion of $r_b(t)$ curve. 
	\item Find $\overline{\Delta T}(r)$ by averaging $\Delta T(r,t)$ over $t$ in the interval $(t_m,t_f)$. Compute $\delta_0(r)$ from Eq.~(10).
	\item Compute $U_b$ as the $r_b(t)$ slope at $t=t_f$.
	\item Use both $\delta_0$ and $U_b$ to re-evaluate $r_c$ thanks to the Landau-Levich equation~(4).
	\item Repeat the steps 2--6 until $r_c$ converges.   
\end{enumerate}
%Since all the variables evolve monotonously in time, 
%The convergence of such an algorithm is quite fast.

\section{Uncertainty analysis}\label{sec:appC}

\begin{figure}[htb]
	\centering\includegraphics[width=0.9\columnwidth,clip]{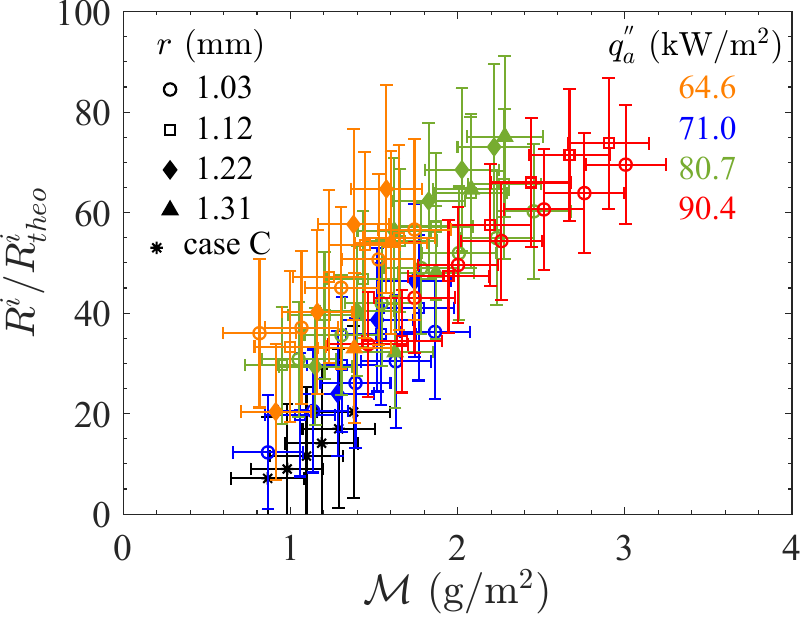} 
	\caption{Same as Fig. 6 of the main article but with the error bars.} \label{fig:AppC:RiMassUncertainties}
\end{figure}
From Eq.~(7) of the main article, the uncertainty of $R^i$ reads
\begin{equation}\label{eq:AppC:DRi}
	\partial R^i = \dfrac{1}{\rho_l \mathcal{L} |\dot\delta|} \partial(\Delta T) + \dfrac{\Delta T}{\rho_l \mathcal{L} \dot\delta^2} \partial(\dot\delta) + \dfrac{1}{k_l}\partial\delta,
\end{equation}
where $\partial f$ denotes the uncertainty of a quantity $f$. The slope $\dot\delta$ is determined by fitting by using many experimental data points. The uncertainty $\partial(\dot\delta)$ is thus determined from the data fit accounting for the uncertainty $\partial\delta/\Delta t$ of one point, where $\Delta t=\SI{0.25}{\milli\second}$ is the sampling time interval and $\partial\delta=\SI{15}{\nano\meter}$ is small thanks to the high accuracy of WLI \cite{JFM24Boiling}. The accuracy of IRT is $\partial(\Delta T)=\SI{0.5}{\kelvin}$ \cite{IJHMT24}. According to Eq.~\eqref{eq:AppC:DRi}, $\partial R^i$ is small for high $|\dot\delta|$, which occurs when either $\Delta T$ (note that its value inside the microlayer area is meaningful) is high or $\delta$ is small, cf. Eq.~(7) of the main article. In the case A \cite{IJHMT24}, $\Delta T\simeq\SI{5}{\kelvin}$ and $\delta\simeq\SI{6}{\micro\meter}$ leading to small values of $|\dot\delta|\simeq\SI{0.25}{\milli\meter/\second}$ and thus to a high uncertainty ($\partial R^i/R^i_{theo}\simeq 40$); for this reason, the case A is not shown in both \autoref{fig:AppC:RiMassUncertainties} and Fig.~6 of the main article. On the other hand, case C \cite{EuroTherm24Exp} showed larger values of $|\dot\delta|\simeq\SI{0.5}{\milli\meter/\second}$ due to a thinner microlayer $\delta\simeq \SI{2}{\micro\meter}$ and $\Delta T\simeq \SI{2}{\kelvin}$. In the case B, $|\dot\delta|$ is also high ($\simeq \SI{0.6}{\milli\meter/\second}$) mostly due to the high wall superheating, the spatio-temporal evolution of which is shown in \autoref{fig:AppC:DT28}. Figure~\ref{fig:AppC:RiMassUncertainties} presents $R^i$ as function of $\mathcal{M}$ with the error bars.

One needs the $\delta_0$ uncertainty to evaluate that of $\mathcal{M}$,
\begin{equation}
	\partial \mathcal{M} = \rho_l (\partial \delta_0+\partial \delta). 
\end{equation}
\bibliographystyle{elsarticle-num-names}
\bibliography{ContactTransf,Taylor_bubbles,PHP,Books}